\documentclass[manuscript]{acmart}

\begin{document}

\title{Exploring AI Peer in Collaborative Mathematics Learning}

\author{Yiran Lin}
\authornote{Corresponding author.}
\email{yiran.lin@mail.mcgill.ca}
\affiliation{%
  \institution{Independent Researcher}
  \country{Canada}
}

\author{Ye Deng}
\affiliation{%
  \institution{University of Pittsburgh}
  \country{USA}}
\email{yed19@pitt.edu}

\author{Ziqi Pan}
\affiliation{%
  \institution{Durham University}
  \country{United Kingdom}
}

\author{Yongan Yu}
\affiliation{%
  \institution{McGill University}
  \country{Canada}
}

\author{Alexandra I. Cristea}
\affiliation{%
  \institution{Durham University}
  \country{United Kingdom}
}

\begin{abstract}
Artificial intelligence (AI) agents are increasingly explored as collaborative learning partners in mathematics education, yet it remains unclear what makes an AI recognizable as a peer. This study examines how domain experts perceive and interact with an adaptive LLM-based peer agent combining a Big Five-informed persona with motivational state modulation for collaborative mathematics problem solving. Through thematic analysis of expert interviews, we identify conditions shaping peer perceptions, including emotional safety, shared control, and collaborative reasoning. Participants primarily judged the AI through its social behavior rather than mathematical capabilities. The AI was perceived more as a peer when it created a non-judgmental space, engaged in tentative reasoning, displayed appropriate imperfection, and respected participants’ pace. Conversely, blind agreement and over-instruction reduced trust and shifted perceptions toward a tutor role. Based on these findings, we propose \textbf{human-centered design dimensions for AI peer agents} to guide future peer-oriented learning systems.

\end{abstract}

\maketitle

\section{Introduction}

Collaborative problem solving is an important part of mathematical learning, where learners explain their reasoning, compare ideas, question one another, and jointly construct solutions \cite{hansen2022students}. Yet educational AI has traditionally been designed not as a collaborator but as an instructor, one focused primarily on diagnosing performance, delivering feedback, and guiding learners toward correct solutions \cite{letourneau2025systematic}. This authority-oriented position differs from that of a peer, who participates in shared reasoning, negotiates meaning, expresses uncertainty, and shares responsibility for the problem-solving process \cite{vygotsky1978mind}. With the advent of large language models (LLMs), educational AI has grown markedly more conversational, inspiring researchers to explore its potential as a learning peer rather than an instructional authority \cite{joo2025ai, lo2025noel}. 

Emerging research illustrates both the promise and the difficulty of this shift. Lyu et al. \cite{lyu2026designing} found that middle-school students valued AI peers that were mathematically competent, while remaining deferential to students' control over the problem-solving process. Research on LLM-based peer agents has similarly explored AI as a participant or moderator in collaborative learning \cite{liu2024peergpt}. Together, these studies suggest that AI peer interaction involves tensions between expertise and deference, assistance and control, and cognitive support and social participation. What remains unclear is how these behaviours collectively shape the social-cognitive role that an AI occupies during collaborative learning \cite{yan2025passive}.

This question is complicated by the fact that role is not simply a design output but also a matter of user interpretation, as users do not always interpret AI social characteristics as intended at design stage \cite{lysova2026chatbots}. Amin et al. \cite{amin2026creating} found substantial variation in how GenAI personas are evaluated and a lack of consistent evaluation practices across the literature. 
For AI peers, understanding this distinction requires examining how learners interpret the agent's behaviour as the interaction develops.

This perspective also raises a design challenge: how can an AI system adapt its behaviour to sustain a peer-like relationship across different moments of learning? Personality and motivation offer complementary mechanisms for shaping such adaptive behaviour. Personality provides relatively enduring tendencies in how a learning partner communicates, responds, and approaches social interaction, while motivation can influence how these tendencies are expressed in response to situational demands \cite{bardach2025motivation}. Recent work on LLM personality similarly shows that personality can be expressed through patterns of language and interaction, while highlighting the limitations of treating personality as a static set of behavioural instructions \cite{li2024bigchat}. In a learning interaction, the relationship between personality and motivation may therefore allow an AI peer to vary its interpersonal behaviour according to the evolving context rather than maintaining a fixed persona. 

To investigate this possibility, we designed an LLM-based agent for collaborative mathematical problem solving that combines personality embeddings based on the \textit{Big Five traits} \cite{zillig2002we} with motivational state modulation. We examine how participants interpret the agent's social role during interaction, with particular attention to moments in which the agent is experienced as more or less peer-like. Rather than evaluating learning gains or mathematical performance, we focus on the interactional conditions that shape perceptions of the AI as a peer.

We address the following research questions:

\begin{itemize}
    \item RQ1: How do participants perceive and interpret the AI's peer role during collaborative mathematical problem solving?
    \item RQ2: What do moments of discomfort, confusion, or resistance reveal about the interactional boundaries of AI as a peer?
\end{itemize}

By examining AI as a peer through its interactional and social-cognitive role, we contribute an empirical investigation of AI-mediated collaborative mathematical learning and propose a \textit{human-centered set of design and evaluation dimensions for AI peer agents}. These dimensions provide a basis for designing and evaluating AI systems intended to participate in learning as peers rather than tutors or authorities.

\section{Related Work}

\subsection{AI-Powered Collaborative Learning in Mathematics}

Collaborative problem solving (CPS) in mathematics involves articulating reasoning, comparing solution paths, and checking ideas with others \cite{hansen2022students}. Research on AI-supported CPS has developed along several different configurations of the agent's role. Early collaborative intelligent tutoring systems incorporated shared problem states, scripted interaction prompts, and representational scaffolds to coordinate group activity \cite{ramachandran2017automated, chen2021instant}. Other systems have used AI to monitor group interaction, identify confusion or disengagement, and provide targeted prompts \cite{troussas2023collaborative}. In these approaches, AI operates around the collaborative activity by organizing, monitoring, or facilitating interaction.

Teachable agents represent a different approach in which learners take responsibility for explaining and evaluating knowledge. Systems such as SimStudent and related learning-by-teaching environments have used artificial agents that request hints, make errors, and learn from learners' feedback and demonstrations \cite{matsuda2013cognitive, matsuda2015teaching}. The emergence of LLMs has further expanded the range of roles that educational agents can occupy by enabling agents to generate responses dynamically and participate in extended dialogue rather than relying on predefined responses or narrowly structured learning activities. Xing et al. \cite{xing2025development} describe ALTER-Math as an effort to develop more naturalistic and personalized teachable-agent interactions in mathematics learning. Building on these capabilities, recent systems have begun to position AI as a collaborator or peer-oriented learning partner \cite{liu2024peergpt, lyu2025exploring}. 

Different roles suggest different design considerations for educational agents. When an agent participates directly in a collaborative activity, its contribution is embedded in an ongoing interaction rather than delivered as an isolated instructional response \cite{joksimovic2023opportunities}. Its timing, conversational behavior, response to uncertainty, and adaptation to the learner can consequently become part of the learning experience. These interactional characteristics provide the context in which personality and motivation can be incorporated into the design of an AI learning peer. They offer a way to model both relatively characteristic patterns of social behavior and the ways these behaviors may change in response to the evolving demands of a learning interaction.

 \subsection{Personality and Motivation}


The interplay of personality and motivation is well established in the field of education. In collaborative learning, both personality and motivation are expressed through interactions and can influence learners' engagement, sense of security, and willingness to participate \cite{fransen2011mediating,bruffee1999collaborative}. 
However, as contemporary integrative perspectives argues, it is important to distinguish relatively enduring personality tendencies from more context-sensitive motivational processes and characteristic adaptations \cite{bardach2025motivation}. This perspective provides a basis for understanding how interpersonal tendencies can be expressed differently as interactional conditions change.

This perspective has increasingly informed the design of educational and conversational agents. In mathematics education, Lyu et al. \cite{lyu2025exploring} examined Big Five personality traits in teachable AI agents and found that different personality configurations were associated with differences in students' cognitive and affective interaction behaviors. HCI research has similarly begun to treat personality expression as an interactional design variable in LLM-based agents. Rahman and Desai \cite{rahman2026vibe} showed that the degree of Big Five personality expression influenced users' perceptions of conversational agents, with medium levels of expression receiving more favorable evaluations than low or high levels. Together, this work suggests that personality in an AI agent is expressed through interaction rather than functioning simply as a fixed persona.

Recent work has consequently begun to consider how personality expression can adapt to context. Fluid Personality Framework proposes adapting both an agent's persona and the intensity of its personality expression according to task context, user goals, user characteristics, and situational demands \cite{rahman2026behavior}. This perspective considers how an agent's social presentation can change as an interaction develops, and such adaptation is consistent with psychological accounts that distinguish enduring personality tendencies from context-sensitive states and characteristic adaptations \cite{bardach2025motivation}.

For an AI learning partner, these perspectives provide a basis for connecting personality expression with motivational processes. Personality can provide relatively enduring interpersonal tendencies, while motivational states can influence how those tendencies are expressed in response to changing goals and interactional conditions \cite{bardach2025motivation}. In collaborative learning, an agent may therefore express encouragement, uncertainty, persistence, or restraint differently as the learner explores an idea, encounters difficulty, or becomes disengaged. Designing these mechanisms together provides a basis for an adaptive AI peer whose social behavior can change with the unfolding interaction rather than remaining tied to a fixed persona.

\section{Adaptive AI Peer Agent Design}
We developed an adaptive AI learning peer designed to simulate a human peer engaged in dialogic problem-solving and mathematical reasoning, by adjusting personality via LLM fine-tuning, and both behavior and interaction styles through structured prompting.

\subsection{Peer Agent Design Overview}

\begin{figure}[htbp]
    \centering
    \includegraphics[width=\textwidth, height =8cm]{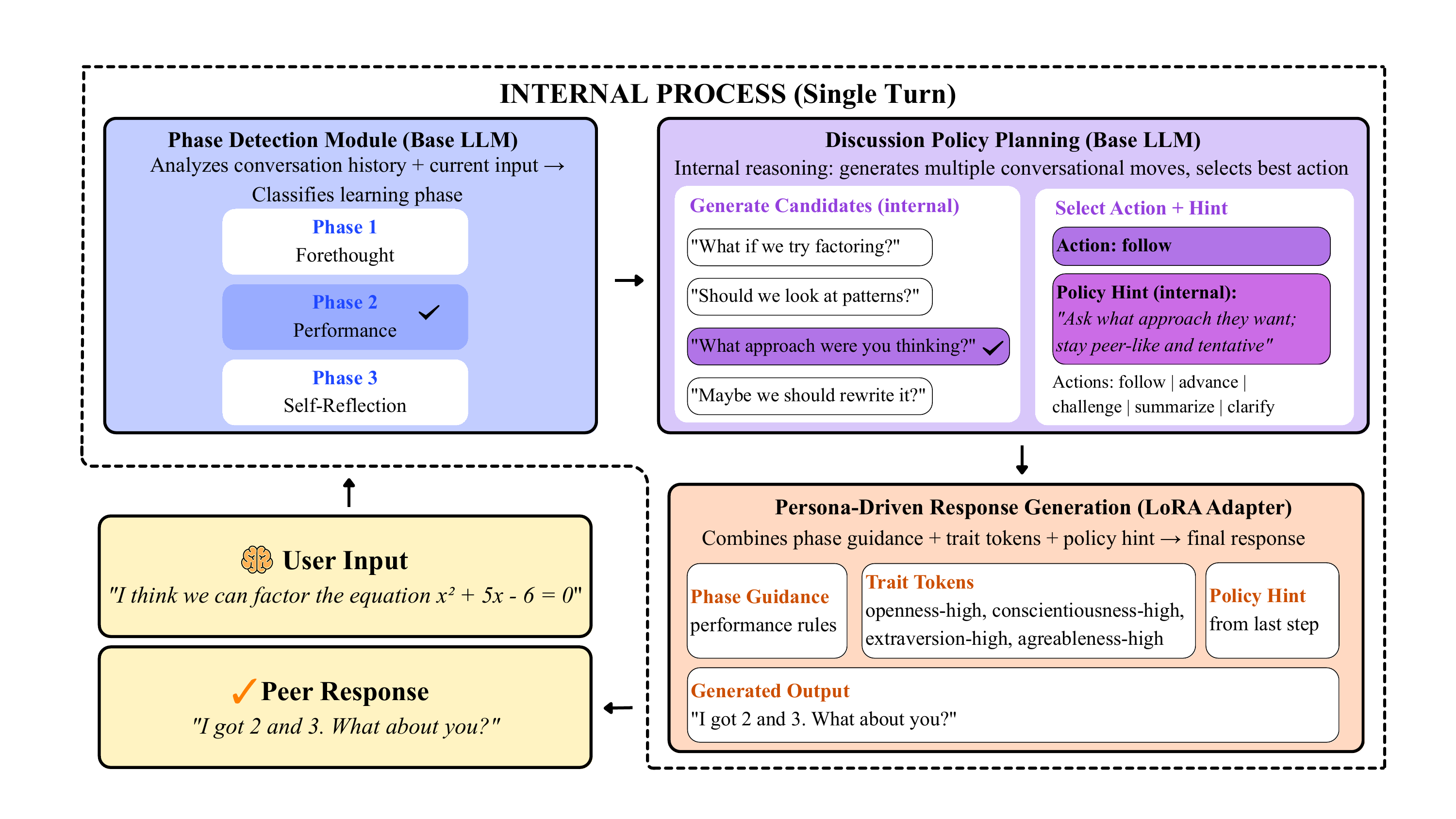}
    \caption{Single-turn interaction flow of the multi-phase Adaptive AI peer Interaction. For each user input, the design performs three steps: (1) phase detection classifies the dialogue state; (2) policy planning selects a peer action type and generates a policy hint; (3) response generation produces a single reply using phase-specific Big Five trait tokens, the policy hint, and global rules that constrain the agent to act as a non-authoritative peer. This process repeats as the dialogue progresses across phases.}
    \label{fig:workflow}
\end{figure}

As shown in Figure \ref{fig:workflow}, the AI peer is designed to adapt its behavior to the evolving interaction. The design is guided by global peer rules and implemented through three coordinated components: (1) phase control, (2) persona embedding, and (3) phase-specific behavioral guidance. 

In each turn, the user input and recent history are sent to an LLM-based phase detector to identify whether the interaction is in the forethought (Phase 1), performance (Phase 2), or self-reflection (Phase 3) \cite{bardach2025motivation}. This phase information conditions both the agent's social stance and its conversational goals. A discussion policy then selects an appropriate type of peer move, and the final response is generated by a persona-conditioned LLM. Together, these components allow the agent to adapt how it participates in the dialogue rather than following a fixed interaction script.

\subsection{Adaptive Features}

\subsubsection{Global Peer Rules}
To ensure peer-like interactions, motivated by prior work on peer learning  \cite{webb2009teacher}, we design global rules that emphasize balanced participation by discouraging evaluative or authoritative positioning, maintaining a collaborative relationship in which responsibility for reasoning is shared. The agent is also prompted to engage the learner in reciprocal dialogue and shared problem ownership, refraining from providing final solutions unless the learner has verified them \cite{kumpulainen2003interpersonal}.

\subsubsection{Phase Control} \paragraph{Phase Classifier Prompt} Phase control is implemented through a deterministic, prompt-based LLM classifier that assigns interactions to three continuous phases labeled \{1,2,3\}. The classifier processes a short window of chat history and determines the current phase and corresponding agent behavior: \textit{forethought} (phase 1) when the task is still being framed, \textit{performance} (phase 2) when active work is underway, and \textit{self-reflection} (phase 3) when progress stalls. This approach follows a process-oriented view of self-regulated learning, in which phases reflect users' engagement states, exploring, performing, or reflecting, rather than performance outcomes \cite{bardach2025motivation}. Grounding phase detection in interaction cues enables the agent to adapt its peer behavior to the user's ongoing thought process, rather than reacting to right or wrong answers.

\paragraph{Discussion Policy as Phase-conditioned Action Selection}
After phase detection, the design applies a discussion policy that determines how the agent should respond as a peer in the next turn. The policy generates several candidate conversational moves and selects one action type: \textit{follow}, \textit{advance}, \textit{challenge}, \textit{summarize}, or \textit{clarify} based on the current interaction phase and recent dialogue context. Phase information directly constrains this choice: exploratory moves are favored in forethought, progress-oriented moves in performance, and reframing or consolidation moves in self-reflection. The selected action is encoded as an internal \textit{policy hint} that guides the final response generated by the persona-tuned model. Following prior work showing that multiple candidate generation reduces rigidity in LLM-based problem solving \cite{asano2025can}, we adapt this idea to dialogue management by selecting among interactional moves rather than mathematical solutions.

\subsubsection{Persona Embedding}\paragraph{Stabilizing Personality with Fine-tuning}Traditional persona design often suffer from being static and abstract \cite{matthews2012designers}. To obtain a stable and recognizable peer persona, we fine-tune a pre-trained LLM (Llama-3.1-8B-Instruct) using Low-Rank Adaptation (LoRA), rather than relying solely on prompt-based persona control. Prior research suggested that textual prompting persona control can struggle to align model responses with persona-specific data \cite{huang2024selective}. Motivated by this observation, we adopt a parameter-efficient fine-tuning approach to encode persona-behavior.

\paragraph{Adaptive Personality Embedding}
To support a stable peer-like interaction style, we encode personality using the Five Factor Model \cite{mccrae1992introduction} and adopt the personality modeling and training framework proposed by Li et al. \cite{li2024bigchat}, enabling the model to learn persona-consistent conversational behavior. Personality expression is adaptive rather than fixed, each phase utilizes a distinct combination of Big Five traits to align with the user's current needs, offering a personalized response through two primary features. The forethought phase emphasizes high \textit{Openness} to support curiosity and exploration; the performance phase combines \textit{Openness}, \textit{Conscientiousness}, \textit{Agreeableness}, and \textit{Extraversion} to support coordinated problem solving; and the self-reflection phase activates high \textit{Conscientiousness} and low \textit{Neuroticism} to support calm review \cite{bardach2025motivation}. Trait tokens act as conditioning signals for the LoRA-adapted model, activating persona-related behavioral tendencies learned during fine-tuning. 

\paragraph{Behavioral Cues and Motivation Injection}

Based on Bardach's framework \cite{bardach2025motivation}, we designed phase-specific behavioral cues that guide how the AI peer should act. In the forethought phase, the peer surfaces intentions without committing to specific methods or narrowing the problem. During performance, it generates ideas, monitors progress, refocuses scattered discussion, and suggests tentative next steps. In the self-reflection phase, it examines factors affecting progress, frames difficulties as task-related, proposes next steps without judgment, and avoids signaling finality.

Second, phase guidance provides the model with specific motivational goals and natural language behaviors. These include ``curiosity-driven" motivation during forethought, "sustained engagement" during the performance phase, and "consolidation" during the self-reflection phase. Together, these mechanisms allow the agent to act as an adaptive peer. By tailoring its personality and communication style to the student's stage in the task, the agent ensures the interaction remains responsive to the specific cognitive demands of the learning process.

\section{Methodology}
\subsection{Pilot Study}
Before the formal study, we conducted a pilot study with two experts in educational technology. The pilot aimed to validate the clarity of the study scenario, the realism of the agent's behavior, and the usability of the design. Participants in the pilot interacted with the agent and provided open-ended feedback on the interaction flow, personality expression, and perceived peer-likeness. Data from the pilot study were not included in the analysis. 

\subsection{Study Procedure}
We adopted an exploratory qualitative approach to examine how the theory-informed AI peer design was experienced and interpreted in interaction \cite{deep2025she}. The initial design was grounded in theories of peer learning and social interaction, and the study aimed to understand how these theoretical principles translated into the agent's interactional behaviors and how they were perceived during use. Rather than testing learning outcomes, we focused on identifying interactional patterns, tensions, and characteristics that could inform the development of a framework for designing AI agents as learning peers.

We conducted the qualitative study between December 2025 and January 2026 with seven researchers in User Experience, Learning Sciences, Educational Technology, and Computer Science, with experience ranging from 3 to over 20 years. Participants were asked to critically engage with the AI agent as a peer and examine how the proposed design dimensions manifested during interaction. In particular, we explored how participants interpreted the agent's social role, perceived its position of authority, and identified moments of collaboration, role shifts, or interactional breakdown. The study was approved by the university ethics board, and informed consent was obtained from all participants.

\begin{figure}[htbp]
    \centering
    \includegraphics[width=0.85\textwidth, height=7cm]{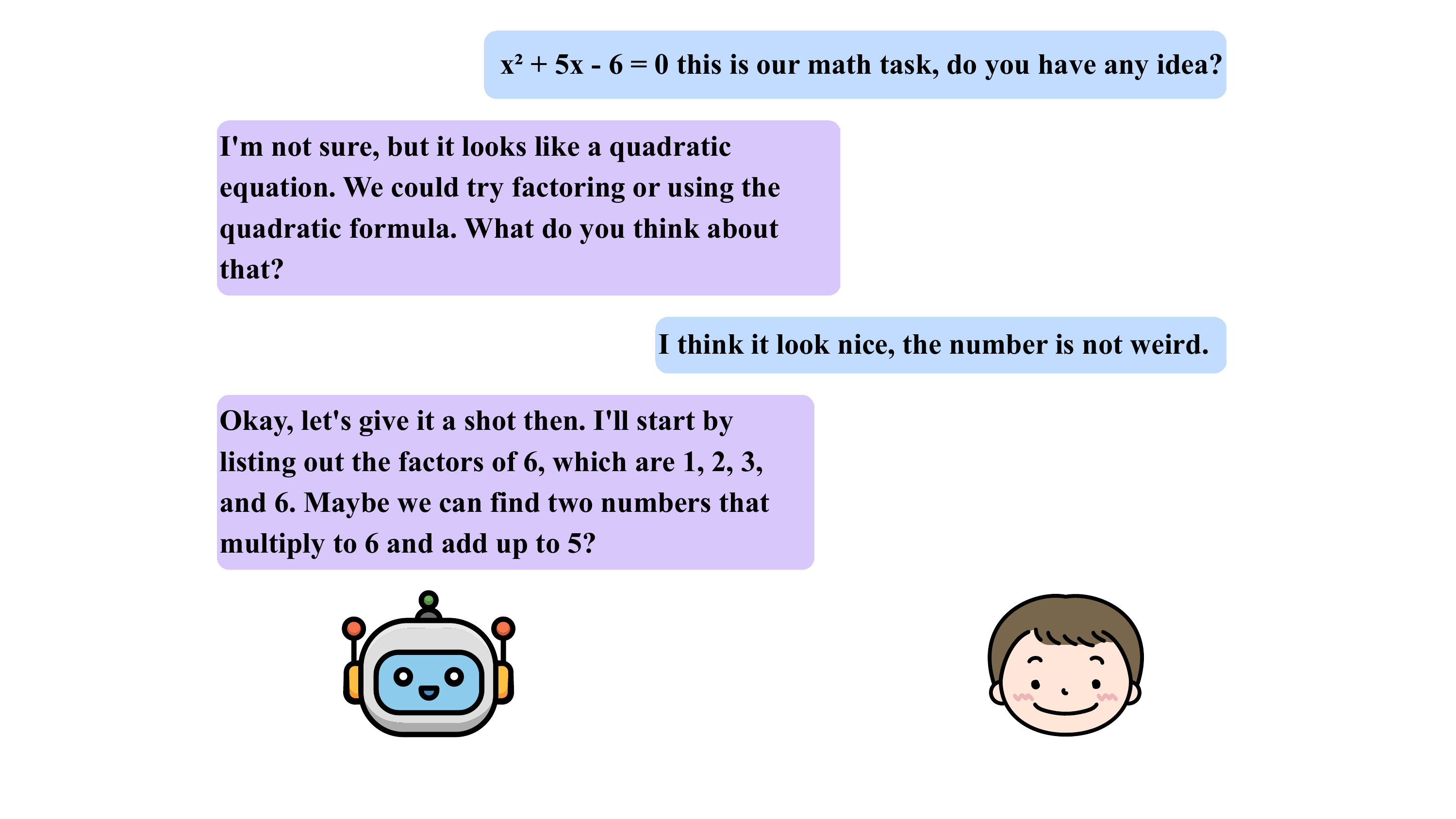}
    \caption{Collaborative Mathematical Problem Solving Example Dialogue during Interview}
    \label{fig:interaction}
\end{figure}

Each session lasted approximately 30 minutes and was conducted through an online meeting platform. After consent, participants were asked to take the perspective of a student learning quadratic functions. Participants then interacted with the AI agent for approximately 20 minutes to collaboratively solve quadratic equations (Figure~\ref{fig:interaction}). They received the same information about quadratic functions that had been provided during the agent's development and were encouraged to treat the agent as a classmate and reason collaboratively, rather than expecting direct instruction or final answers. No specific conversational structure was imposed so as to preserve the natural development of the interaction.

Following the interaction, participants completed a structured interview based on predefined questions derived from the research questions. The interview focused on how they interpreted the agent's social role and authority, how the interaction reflected the proposed design dimensions, and where they observed moments of collaboration, role shifts, tension, or breakdown.

\subsection{Data Analysis}
All sessions were audio-recorded with participants' permission and transcribed. The interview transcripts were de-identified before analysis. We conducted an iterative thematic analysis \cite{terry2017thematic}. Two authors independently reviewed the transcripts, generated initial codes, and then collaboratively grouped them into themes, through discussion and refinement.

\section{Results}
\subsection{RQ1: Perception of an AI peer agent} 

\subsubsection{Reduced Pressure and Increased Safety}
A dominant theme across the interviews was that participants associated a stronger sense of safety with interacting with the AI compared to a real peer, when considering the experience from a learner's perspective. This allowed them to focus on the math problem without worrying about being judged. P1 explained that when compared to regular class setting, seeing other students finish faster made them feel ``anxious and ashamed," while with the AI there was no comparison. P3 shared that because the agent was “not a real human being," they felt ``less embarrassed" and ``more comfortable making mistakes"; P5 also emphasized that safety came from the lack of human judgment: ``I know he won't think I'm fool...He doesn't have any human emotions, he's just AI."
P6 further explained that safety increased when the AI felt like an equal. They suggested that if the agent was presented as ''your classmate...same level as you," then ``it feels safer", as ``with your friend or someone equal to you, you can make mistakes without a problem." This shows that safety was created both by the AI's non-human nature and by the feeling of equality. For these participants, experiencing the AI as a peer meant being able to make mistakes without fear of shame.

\subsubsection{Peer Through Imperfection}
Participants described the AI's peer-likeness through its willingness to acknowledge its own imperfections, which are caused by the inherent imperfections of LLM. P3 felt the agent was peer-like because it was ``not very clever" and made ``stupid and simple mistakes." When P3 gave a wrong answer and the agent accepted it, they said it felt “like a real human peer." P7 expressed a similar view; the agent felt like a peer, as it ``did not show the solution directly," but instead ``made some assumptions and asked for my idea." They also noted that the agent ``made mistakes sometimes" and tried wrong approaches before asking for participants' opinion, which made the interaction feel collaborative.  

However, not all participants welcomed this behavior. P5 disliked an AI that was ``as stupid as I am," and believed a peer should be ``smarter than me" to provide helpful ideas. This shows that while mistakes helped some participants see the AI as human-like, others felt that too many mistakes reduced its usefulness. This indicates a distinction between peer-likeness and peer-usefulness, which were not always aligned in participants' experiences.

\subsubsection{The Impact of Personality and Motivation Embeddings}
Personality and motivation had different effects on participants. P1 felt the agent's personality was ``steady" and that it ``really wants me to learn," which made the interaction feel meaningful. However, P7 did not experience the personality as stable. They said it was ``more a surface level" and changed according to the user’s input. They added that they mainly saw the AI “as a tool, not ... with personality… like friends." This shows that for some participants, the personality design was not strong enough.  
Personality also caused problems when the tone felt inappropriate. P1 became anxious when the agent said it was ``feeling a bit unconcerned," which sounded dismissive. Others described the agent as ``strict" (P4) or a ``know-it-all" (P2). When the agent sounded confident or directive, participants were more likely to see it as a teacher rather than a peer. This shows that personality can help create consistency, but it can also create distance, if the tone is too authoritative.

\subsubsection{Collaboration versus Instruction}
Participants described collaboration as situations where the AI guided their thinking without taking control. P6  felt like ``working together as a team" because the agent ``has the information but  … gives you hints and lets you think and be part of the solution."

This feeling changed when the AI moved too quickly. P7 felt ``a little annoyed" when they were still thinking and the agent ``just gave me the right number." For them, showing the answer too early broke the sense of collaboration. Similarly, P4 felt the agent was ``examining me, not helping me," and P2 said it was ``very fixed on a direction of travel." In these moments, the AI was no longer seen as a peer, but as a tutor giving instructions. In summary, participants experienced the AI as a peer when it supported their thinking process, and as a teacher when it controlled the pace or direction of the solution.

\subsection{RQ2: Boundaries of AI as a learning peer}

\subsubsection{Failure of Shared Memory and Mutual Understanding}
The most significant boundary emerged through moments of confusion, where the AI failed to maintain a consistent shared understanding. For a peer relationship to be effective, both partners must remember what was previously discussed. P2 expressed frustration when the agent ``claims I've said some things which I didn’t," which led to a direct ``breakdown in trust". P3 experienced a similar issue at the very start of the conversation, when the agent assumed they were already solving an equation that hadn't been introduced yet. P3 remarked, ``I don't remember anything about giving an equation... its previous memory might prevent our mutual understanding." These point to a boundary that the AI cannot yet act as a reliable, long-term peer, because it often hallucinates or mixes up different conversations.

\subsubsection{Blind Validation and Misinformation}
Another boundary appeared through discomfort when the AI agreed with incorrect ideas. P4 asked, "Is he agreeing because it's correct or because he doesn’t know the correct answer?" P2 warned that this could “keep reinforcing those mistakes forever.”  P7 believed this issue came from the design rather than from the idea of an AI peer, ``I would rather say it's due to the design... if it can hold on and not show the correct answer.” This shows that participants accepted the idea of an AI peer, but expected better control over how and when correctness is provided.

\subsubsection{Lack of Flexibility}
Resistance emerged when the AI enforced a fixed problem-solving order. P5 said the agent ``doesn't want me to jump the steps." P2 felt it did not ``really try and collaborate" with their ideas. P7 also suggested that the agent should wait longer before giving answers. Overall, AI was seen as less peer-like when it followed a rigid process, instead of adapting to the participants' approach.

\subsubsection{Emotional and Social Gaps}
Finally, participants noted that the AI could not respond well to their emotions. P1 said the agent could not ``sense your tone, your emotions or motivations," and sometimes sounded “harsh or abrupt.” When the agent said it was unconcerned,” P1 felt more stressed.  

P6 believed that younger students would feel comfortable using AI and said the design was “good,” but still suggested improvements, such as making the system faster and smoother. Overall, most discomfort came from limits in design, such as timing, pacing, memory, and emotional awareness, rather than from rejecting the idea of AI acting as a peer.

\section{Discussion and Design Implications}

This study contributes to emerging work on AI-supported collaborative learning by examining what it means for an AI system to participate as a peer rather than as an instructional authority. Our findings suggest that the peer role is not a fixed property of an agent, but an interactional accomplishment. Participants continually interpreted the agent's social position through how it contributed to reasoning, responded to uncertainty, respected their control, and adapted to the pace of the interaction. In this sense, the relevant question for AI peer design is not simply whether the system can perform as a peer, but whether its behavior is experienced as participation in a shared activity.

This perspective offers a theoretical contribution to how AI peers can be understood within collaborative learning. From the sociocultural perspective, learning occurs through shared activity, dialogue, and negotiation rather than through knowledge transmission alone \cite{vygotsky1978mind,dillenbourg1999collaborative} and this distinction remains important when one of the participants is an AI system. Mathematical competence was valued, but competence became tutor-like when the agent assumed control, disclosed solutions prematurely, or moved ahead of the learner's reasoning. Conversely, tentative contributions and space for learner reasoning could support a stronger sense of joint problem solving, even when they were less efficient. This helps explain why \textit{peer-likeness} and \textit{usefulness} should not be treated as interchangeable outcomes. 
Our findings therefore position shared control, rather than competence alone, as an important relational property of AI-supported collaboration.

The study also contributes empirically by showing how social and epistemic qualities interact in participants' experiences of an AI peer. Participants described reduced social pressure and greater comfort expressing uncertainty, consistent with research showing that evaluative concerns and fear of error can constrain participation in mathematics \cite{beilock2015math,fransen2011mediating}. Yet participants did not interpret all forms of uncertainty as peer-like. Tentative reasoning could signal shared exploration, whereas blind agreement or internally inconsistent responses weakened the agent's epistemic credibility. Productive peer collaboration similarly depends on participants contributing ideas that can be questioned, revised, and incorporated into shared reasoning \cite{barron2003smart,mercier2011humans}. Thus, reducing social evaluation can create room for exploration, but the agent still needs to contribute reasoning that participants can treat as meaningful.

Our findings further connect AI peer design with emerging HCI work on personality and adaptive agent behavior. Research in mathematics education has shown that Big Five personality configurations can shape students' cognitive and affective interactions with teachable AI agents \cite{lyu2025exploring}, while recent HCI studies suggest that the degree of personality expression and its adaptation to context can influence how users experience LLM-based agents \cite{rahman2026vibe,rahman2026behavior}. Our observations provide a learning-context interpretation of this work: behaviors such as encouragement, agreement, uncertainty, persistence, and restraint were interpreted differently depending on the state of the problem-solving interaction. This is consistent with theoretical accounts distinguishing relatively stable personality tendencies from context-sensitive motivational processes \cite{bardach2025motivation}. Our system translates this distinction into the design of an adaptive AI agent: personality shapes its general interaction style, while motivational states allow its behavior to adapt as the learning interaction unfolds.

Finally, the moments in which the interaction broke down provide an empirical basis for translating these findings into design guidance. Failures in shared context, premature answers, rigid sequences, and poor responsiveness did not simply reduce system usability; they could change how participants understood the relationship between themselves and the agent. These observations complement AI-peer research emphasizing pacing, turn-taking, responsiveness, and student control \cite{lyu2026designing}, while also reflecting broader challenges in LLM-based interaction. Based on these recurring interactional conditions, we propose six human-centered design dimensions for AI peer agents: \textit{Emotional Safety, Shared Control, Authority Positioning, Balanced Uncertainty, Memory Consistency,} and \textit{Responsiveness to Pace}, see Table~\ref{tab:ai_peer_standards}. These dimensions are intended as a design-oriented vocabulary rather than a validated evaluation instrument. They provide a way to connect the social-cognitive role of the AI with interactional behaviors that can be considered when designing future peer-oriented learning systems.

\begin{table*}[t]
\caption{Human-centered design dimensions for AI agents intended to function as learning peers}
\label{tab:ai_peer_standards}
\centering
\begin{tabular*}{\textwidth}{@{\extracolsep{\fill}}p{2.6cm}p{6.0cm}p{6.0cm}@{}}
\toprule
\textbf{Dimension} & \textbf{Description} & \textbf{Design Implication} \\
\midrule

\textit{Emotional Safety} &
Learners feel able to express uncertainty and make mistakes without fear of evaluation &
Avoid evaluative language; normalize hesitation and partial reasoning \\

\textit{Shared Control} &
Problem solving is jointly constructed and learner agency is preserved &
Avoid enforcing fixed sequences; allow learners to set pace and direction \\

\textit{Authority Positioning} &
The agent is experienced as a social equal rather than an instructor &
Use tentative language; avoid directive or corrective tone \\

\textit{Balanced Uncertainty} &
The agent expresses hesitation while maintaining internal coherence &
Signal doubt without default agreement; pause when reasoning seems inconsistent \\

\textit{Memory Consistency} &
The interaction reflects a stable shared history &
Avoid hallucinated references; maintain local conversational coherence \\

\textit{Responsiveness to Pace} &
The agent adapts to learners' timing and engagement &
Delay answers during exploration; slow down when confusion emerges \\

\bottomrule
\end{tabular*}
\end{table*}

The dimensions capture complementary aspects of peer interaction. \textit{Emotional Safety}, \textit{Shared Control}, and \textit{Authority Positioning} describe the social relationship between the learner and the agent. \textit{Balanced Uncertainty} concerns how the agent communicates its epistemic position, while \textit{Memory Consistency} and \textit{Responsiveness to Pace} concern continuity and timing across the interaction. Together, they shift attention from whether an AI can perform a learning task toward how its participation is experienced within that task.

The dimension also distinguishes \textit{peer-likeness} from \textit{usefulness} as related but separate design goals. An agent can be useful because it provides efficient explanations or advances a solution, while the same behavior can reduce the learner's sense of shared control. Conversely, behaviors that preserve peer-like interaction may sometimes require slower or less directive responses. Future work can examine whether these dimensions generalize across learners, domains, and agent architectures, and whether they can be operationalized as evaluation criteria for AI peer systems.

\section{Limitations and Future Work}

This study has several limitations; first, our analysis focused on a single AI peer prototype. The agent combined Big Five personality embeddings with motivational state modulation, and some interactional patterns may therefore reflect specific choices in the agent architecture, prompting, and adaptation mechanisms. The proposed framework should consequently be understood as a theory-informed set of design dimensions rather than a validated evaluation instrument. Comparative studies across different agent architectures, personality configurations, and adaptation strategies are needed to examine how consistently the proposed dimensions apply across AI peer systems.

Second, the current study examined expert interpretations of the AI peer rather than students' lived experiences of using the system. This distinction was intentional given our focus on identifying design considerations and interactional boundaries before evaluating educational effectiveness. A subsequent phase should examine how students experience the same interactional dimensions during actual collaborative mathematical problem solving, including whether the conditions identified by experts correspond to students' perceptions of emotional safety, shared control, uncertainty, and peer-like interaction. Such studies could further investigate how these experiences relate to engagement, collaborative processes, and learning outcomes.

\section{Conclusion}

This study responds to the growing attention in AIED that looks beyond designing AI systems merely as instructional tools. We examined how learners experience an AI agent in the social–cognitive role of a peer, laying the groundwork for future research on AI systems that emulate human-like social and cognitive behaviors in learning environments. Our findings show that in peer-oriented designs, correctness alone is insufficient; the enactment of authority, uncertainty, and shared responsibility is central.

We contribute a human-centered set of design and evaluation dimensions for AI peer agents that emphasize interaction quality, emotional safety, and shared control. This perspective clarifies AI peers not as simplified tutors or information providers, but as partners that support collaborative reasoning and sustained engagement.

\bibliographystyle{ACM-Reference-Format}
\bibliography{acmart-primary/bib/sample-base}

\end{document}